\documentclass{article}
\usepackage{graphicx}

\makeindex
\begin{document}

\title{\bf A New Experiment to Study Hyperon CP Violation, Charmonium, and Charm\footnotemark
}
\author{Daniel M. Kaplan\footnotemark\\[0.1in]
\it Physics Division, Illinois Institute of Technology\\
\it Chicago, Illinois 60616, USA}

%\maketitle

\twocolumn
[\begin{small}\maketitle\abstract{
Fermilab operates  the world's most intense antiproton source, now exclusively dedicated to serving the needs of the Tevatron Collider. 
The anticipated 2009 shutdown of the Tevatron presents the opportunity for world-leading low- and medium-energy antiproton programs. We summarize the  status of the Fermilab antiproton facility and review physics topics for which a future experiment could make the world's best measurements.
}\vspace{0.4in}\end{small}]

\renewcommand{\thefootnote}{\fnsymbol{footnote}}
\footnotetext[1]{Contributed to XXIII International Symposium
on Lepton and Photon Interactions at High Energy (LP07), Daegu, Korea, Aug.\ 13--18, 2007.} 
\footnotetext[2]{E-mail address: kaplan@iit.edu}
\renewcommand{\thefootnote}{\arabic{footnote}}
\section{Overview}

Fermilab operates the world's highest-energy (8\,GeV kinetic) and highest-intensity antiproton source. The luminosity needs of the Tevatron Collider have engendered a continuous performance-improvement program, so that the  stacking rate, %$\approx$\,20\,mA/hr (or 
$\approx$\,$2\times10^{11}\,{\overline p}$/hr, is now some five times that in E835~\cite{E835-NIM} (the last time the Antiproton Source was used for medium-energy experiments), and an order of magnitude beyond that planned~\cite{Kramer-PAC07} for the  Gesellschaft f\"ur Schwerionenforschung (GSI) Facility for Antiproton and Ion Research (FAIR) in Darmstadt, Germany~\cite{FAIR}. With the planned 2009 shutdown of the Tevatron, the Fermilab Antiproton Source could once again become available for medium-energy experiments.

Using the Antiproton Source, Fermilab experiments E760 and E835 made the world's most precise measurements of charmonium masses and widths~\cite{E835-NIM,E835-psi-widths}. This precision ($\stackrel{<}{_\sim}$\,100\,keV) reflects the narrow energy spread of the stochastically cooled antiproton beam and the absence of Fermi motion and negligible energy loss in hydrogen cluster-jet targets. The other key advantage of ${\overline p}p$  annihilation is its ability to produce charmonium states of all quantum numbers, whereas $e^+e^-$ machines produce primarily $1^{--}$ states. 

\begin{figure}
\centerline{\includegraphics[width=\linewidth,bb=52 210 521 660,clip]{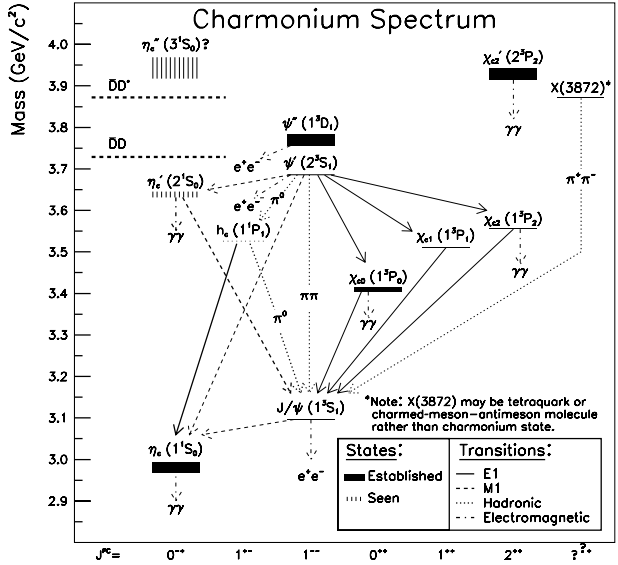}}
\caption{Spectrum of the charmonium system. Shown are masses, widths (or for those not yet measured, 90\% confidence level upper limits on widths), and quantum numbers of observed charmonium states, with some of the important transitions also indicated~\protect\cite{PDG2006,ELQ}.}
\label{fig:ccbar}
\end{figure}

Additional running in the charmonium region would be valuable for clarifying some still-elusive aspects of the charmonium system (Figure~\ref{fig:ccbar}), including the $h_c$ mass and width, $\chi_c$ radiative-decay angular distributions, and ${\eta^\prime_c}(2S)$ full and radiative widths. It would also afford the opportunity for precision studies of a number of recently observed states in the charmonium region whose nature has not been determined: the  $X(3872)$, $X(3940)$, $Y(3940)$, $Y(4260)$, and $Z(3930)$~\cite{ELQ}. As we will see, improved sensitivity is possible not only by virtue of longer running time, but also via higher luminosity and use of a magnetic spectrometer (in contrast to that of E760 and E835, which relied primarily on electromagnetic calorimetry to ``dig" the rare charmonium signals out of the $\sim$\,100\,mb total cross section).

Antiproton annihilation has also proved valuable for hyperon studies~\cite{Hamann}. Possibly the highest-impact issue in hyperon physics today is whether and to what extent hyperon decays violate {\em CP} symmetry. As Table~\ref{tab:HCP} indicates, until 2000 the world's most sensitive search for hyperon {\em CP} violation was by PS185 at LEAR, using an antiproton flux $<10^6$\,Hz. The orders-of-magnitude-higher rate at Fermilab can enable the world's most sensitive search and possibly find so-far elusive contributions due to new physics.

Currently the world's most sensitive hyperon experiment is HyperCP, where a surprise---the observation of apparent flavor-changing neutral currents (FCNC) in hyperon decay~\cite{Park-Sigpmumu}---deserves further experimental attention. 

The rate of $D$-pair production has been estimated at about 100/s for $\sqrt{s}$ near the $\psi(4040)$~\cite{PANDA-TPR}. This could lead to a sample of $\sim10^9$\,events/year produced and  $\sim10^8$/year reconstructed, roughly an order of magnitude beyond the statistics accumulated by the B Factories so far. There is thus the potential for  competitive measurements, e.g., of $D^0$ mixing and possible {\em CP} violation in charm decay.

Table~\ref{tab:thresh} lists  center-of-mass energies and lab-frame antiproton momenta for some processes of possible interest.
%\subsection{\it Our proposal} 
%We are proposing a focused experimental program aimed at those measurements for which the Antiproton Source is best suited: (1) precision studies of states in the charmonium region and (2) the search for new physics in hyperon decay. 
The measurements mentioned above can 
be performed with a common apparatus using existing technologies. Depending on available resources, existing detector components might be recycled for these purposes; alternatively, modest expenditures for new equipment could yield improved performance. %The opportunity for such studies will soon arrive, with the planned 2009 shutdown of the Tevatron.  The importance of these measurements justifies the resumption of such a program at Fermilab.
%is now $\approx$\,20\,mA/hr (or $2\times10^{11}\,{\overline p}$/hr), five times that in E835.\cite{E835-NIM} 
We propose to run with up to ten times the typical E835 luminosity~\cite{E835-NIM} (${\cal L}\stackrel{<}{_\sim}2\times10^{32}\,{\rm cm}^{-2}{\rm s}^{-1}$), via increased store intensity or  target density. %Since stochastic cooling works best with small stacks, more intense stores seem nonoptimal. The E835 cluster-jet target (an upgrade of the E760 one) produced\cite{E835-NIM} up to $\approx2.5\times10^{14}\,$atoms/cm$^2$. Higher cluster-jet density is proposed for the PANDA program (also planned for ${\cal L}=2\times10^{32}\,{\rm cm}^{-2}{\rm s}^{-1}$).\cite{PANDA-TPR} Other options include a plastic or metal wire or pellet in the beam halo,\cite{HERA-B} a solid-H$_2$ target on the tip of a cold finger, or a stream of H$_2$ pellets. A non-H$_2$ target, while suitable for hyperon running,  would destroy the superb energy resolution needed for the charmonium studies. We favor simultaneous charmonium and hyperon running with an H$_2$ target.

\begin{table}
\caption{Thresholds for some processes of interest and lab-frame $\overline{p}$ momentum for  $\overline{p}p$ fixed-target.}
\label{tab:thresh}
\begin{center}
\begin{tabular}{lcc}
\hline\hline
 & \multicolumn{2}{c}{Threshold} \\
Process & $\sqrt{s}$ & $p_{\overline p}$\\
 & (GeV) &  (GeV/$c$) \\
\hline\hline
$\overline{p}p\to\overline{\Lambda}\Lambda$  & 2.231 & 1.437\\
$\overline{p}p\to\overline{\Sigma}{}^-\Sigma^+$  & 2.379 & 1.854\\
$\overline{p}p\to\overline{\Xi}{}^+\Xi^-$ & 2.642 & 2.620 \\
$\overline{p}p\to\overline{\Omega}{}^+\Omega^-$ & 3.345 & 4.938\\
\hline
$\overline{p}p\to\eta_c$ & 2.980 & 3.678\\
$\overline{p}p\to\psi(3770)$ & 3.771 & 6.572  \\
$\overline{p}p\to X(3872)$ & 3.871 & 6.991  \\
$\overline{p}p\to X {\rm \,or\,} Y(3940)$ & 3.940 & 7.277  \\
$\overline{p}p\to Y(4260)$ & 4.260 & 8.685  \\
\hline\hline
\end{tabular}
\end{center}
\end{table}

Further detail on our proposed experimental program may be found below and in \cite{pbar-LoI} and \cite{Kaplan-CTP}.

\section{Physics Examples}

We next consider  representative physics examples: studying the $X(3872)$, improved measurement of the parameters of the $h_c$, searching for hyperon $C\!P$ violation, and studying a recently discovered rare hyperon-decay mode. (This list is not exhaustive; see Sec.~\ref{addl-phys} for additional topics.)

\subsection{$X(3872)$}
The best established of the new states, the $X(3872)$ was discovered~\cite{Belle-3872} in 2003 by the Belle Collaboration via $B^\pm\to K^\pm X(3872)$, $X(3872)\to \pi^+\pi^-J/\psi$, and quickly confirmed by CDF~\cite{CDF-3872}, D\O~\cite{D0-3872}, and BaBar~\cite{BaBar-3872}. %Now seen (Table~\ref{tab:X3872}) in  $\gamma J/\psi$,\cite{Belle-Jpsi-gamma} $\pi^+\pi^-\pi^0J/\psi$,\cite{Belle-3piJ} and $D^0{\overline D}{}^0\pi^0$ modes\cite{Belle-DDpi} as well, i
It does not appear to fit within the charmonium spectrum~\cite{ELQ}. %Although well above open-charm threshold, its  width\cite{PDG2006} ($<2.3$\,MeV at 90\% C.L.) implies that decays to $D{\overline D}$ are forbidden, suggesting unnatural parity,\cite{QWG-Yellow} $P=(-1)^{J+1}$. It is a poor candidate for  $\psi_2\,(1\,^3D_2)$ or $\psi_3\,(1\,^3D_3)$\cite{ELQ,Belle-3piJ,QWG-Yellow} due to nonobservation of radiative transitions to $\chi_c$. The observation of $X (3872) \to \gamma J/\psi$ implies positive $C$-parity, and additional observations essentially rule out all possibilities other than $J^{PC}=1^{++}$.\cite{Belle-LP2005,Braaten-HQW} The  available charmonium assignment with those quantum numbers,  $\chi^\prime_{c1}\,(2\,^3P_1)$, is highly disfavored\cite{ELQ,QWG-Yellow} by the observed rate of $X (3872) \to \gamma J/\psi$. Moreover, the plausible identification of $Z(3930)$ as the $\chi^\prime_{c2}\,(2\,^3P_2)$  suggests\cite{ELQ} that the $2\,^3P_1$  should lie some 49\,MeV/$c^2$ higher than the observed\cite{PDG2006} $m_X=3871.2\pm0.5\,$MeV/$c^2$.
The coincidence of the $X(3872)$ with  $D^0 {\overline D}{}^{*0}$  threshold suggests possible novel interpretations: 
%various solutions to this puzzle, including 
an $S$-wave cusp~\cite{Bugg}, a tetraquark state~\cite{tetraquark}, or a meson-antimeson molecule---a bound state of $D^0 {\overline D}{}^{*0}+D^{*0} {\overline D}{}^0$\cite{molecule}.\footnote{The mass coincidence may be accidental, and the $X(3872)$ a $c{\bar c}$-gluon hybrid state; however, the mass and quantum numbers make it a poor match to lattice-QCD predictions for such states~\protect\cite{ELQ}.} A key measurement is the precise mass difference between the $X$ and that threshold, which should be  slightly negative, in accord with the small molecular binding energy~\cite{Braaten-HQW}:
\begin{equation}0<E_X=(m_{D^0}+m_{D^{*0}}-m_X)c^2\ll10\,{\rm MeV}\,.\end{equation}
(A measurement of the width is also highly desirable.)
Current measurements~\cite{CLEO-mD} give $E_X=0.6\pm0.6$\,MeV/$c^2$, with the uncertainty dominated by that of $m_X$. The ${\overline p}{p}$ formation technique should be able to tighten the uncertainty by nearly an order of magnitude.
 %When our precision measurement is made, it will still  dominate, assuming the total uncertainty on $m_{D^0}$ improves roughly as $1/\sqrt{N}$ as the statistics of the CLEO analyzed sample increase by an order of magnitude.\cite{Rosner-private} 
Additional measurements, including ${\cal B}[X(3872)\to \pi^0\pi^0J/\psi]$  and ${\cal B}[X(3872)\to \gamma\psi^\prime]$, will also contribute to quantum-number determination~\cite{Barnes-Godfrey,ELQ}.

%\subsubsection{$X(3872)$ sensitivity estimate}

The ${\overline p}p\to X(3872)$ cross section  is unmeasured but  estimated to be similar in magnitude to those for $\chi_c$~\cite{Braaten-X-production}. This estimate is supported by the observed rates and  distributions of ${\overline p}p\to X(3872)\,+$\,anything at the Tevatron~\cite{D0-3872} and of $B^\pm\to K^\pm X(3872)$~\cite{PDG2006}, which resemble those for charmonium states. Extrapolation from E760's  $\chi_{c1},\, \chi_{c2}\to\gamma J/\psi$ signals~\cite{E760-chi_c}
%(branching ratios of 36\% and 20\%, respectively\cite{PDG2006}) with  $44\pm2$\% acceptance $\times$ efficiency and $\approx$\,500 observed events  per ${\rm pb}^{-1}$ at each resonance.\cite{E760-chi_c} At $10^{32}\,{\rm cm}^{-2}{\rm s}^{-1}$, the 90\%-C.L.  limit\cite{BaBar-BR} ${\cal B}[X(3872)\to\pi^+\pi^-J/\psi]>0.042$ then 
implies $\stackrel{>}{_\sim}$\,4\,$\times10^3$ $X(3872)$ events per nominal month ($1.0\times10^6$\,s) of running, a rate competitive even with that of the proposed SuperKEKB upgrade~\cite{SuperKEKB} (should that project go forward).

Given the uncertainties in the cross section and branching ratios~\cite{BaBar-BR}, the above may well be an under- or overestimate of the ${\overline p}{p}$ formation and observation rates, perhaps by as much as an order of magnitude. Nevertheless, it appears that a new experiment at the Antiproton Source could obtain the world's largest clean samples of $X(3872)$, in perhaps as little as a month of running. The  high statistics, event cleanliness, and unique precision available in the ${\overline p}p$ formation technique could enable the world's smallest systematics. Such an experiment could thus provide a definitive test of the nature of the $X(3872)$.

\subsection{\boldmath $h_c$}

Observing the $h_c$\,$(1 {}^1P_1)$ charmonium state and measuring its parameters were high-priority goals of E760, E835, and their predecessor experiment, CERN R704. As a narrow state with suppressed couplings both to $e^+e^-$ and to the states that are easily produced in $e^+e^-$ annihilation, the $h_c$ is a difficult state to study experimentally.

A key prediction of QCD and perturbation theory is that the charmonium spin-zero hyperfine splitting, as measured by the mass difference $\Delta m_{\rm hf}$ between the $h_c$ and the spin-weighted average of the $\chi_c$ states, should be close to zero~\cite{HF-pred}. Current PDG-average values~\cite{PDG2006} give $\Delta m_{\rm hf}=-0.57\pm0.28$\,MeV, nonzero at 2$\sigma$ but within the QCD expected range. 

%As shown in Fig.~\ref{fig:hc-ideogram}, t
The PDG-average  $m(h_c)$ value is based on claimed observations by CERN R704 (Baglin {\it et al.})\ of $h_c\to J/\psi\, X$ (5 events)~\cite{Baglin}, E760 of $J/\psi\,\pi^0$ (59 events)~\cite{E760-h_c}, and E835 (13 events) and CLEO (168\,$\pm$\,40 events) of $\eta_c\gamma$~\cite{E835-h_c,CLEO-h_c}. The PDG error on $m(h_c)$ includes a scale factor of 1.5 due to the tension among these measurements. Moreover, the two most precise (E760 and E835) are based on statistically marginal ($<$\,3$\sigma$)  signals, and the reliability of the  E760 result is called into question by the negative results of the E835 $h_c\to J/\psi\, \pi^0$ search~\cite{E835-h_c}. The R704 result is on even weaker ground: a ${\overline p}p\to h_c\to J/\psi\, X$ signal at the level implied by Baglin {\it et al.}~\cite{Baglin} is most likely ruled out by  E760~\cite{Cester-Rapidis} %(as discussed above) 
as well as by E835~\cite{E835-h_c}. 

Thus of the four results used by the PDG, only one is clearly reliable, and the claimed precision on $m(h_c)$ is far from established. This motivates an improved experimental search. Also of  interest are the width and branching ratios of the $h_c$, for which QCD makes clear predictions; the decay modes also bear on the question of isospin conservation in such decays.

E835's  $h_c\to\eta_c\gamma\to(\gamma\gamma)\gamma$ sensitivity was limited by the $(2.8\pm0.9)\times10^{-4}$  $\eta_c\to\gamma\gamma$ branching ratio, and their acceptance $\times$ efficiency was  only $\approx$\,3\% due to cuts against the substantial $\pi^0$ background~\cite{E835-h_c}. With a magnetic spectrometer, likely $\eta_c$ modes include $\phi\phi$, $\phi K^+K^-$, $K^*K^*$,  and $\eta^\prime\pi^+\pi^-$. These have branching ratios up to two orders of magnitude larger, as well as more-distinctive decay kinematics than $\gamma\gamma$, probably allowing looser cuts and thus higher efficiency. For example, the $\phi\phi\to K^+K^-K^+K^-$ final state has no quarks in common with the initial ${\overline p}p$ state and so should contain little background. E835 searched for $\eta_c\to\phi\phi$  but without a magnet it was barely feasible. Assessing the degree of improvement will require detailed simulation work, but at least an order of magnitude in statistics seems likely. Additional improvement will come from the higher luminosity we propose.

Provided detailed simulation studies bear out these ideas, we will soon have the opportunity to resolve this  20-year-old experimental controversy.

\subsection{Hyperon {\boldmath $C\!P$} violation}

%Besides the well-known  {\em CP} violation in $K$- and $B$-meson mixing and decay~\cite{PDG2006} 
The standard model (SM) predicts only slight ($\stackrel{<}{_\sim}$\,$10^{-5}$) hyperon-decay {\em CP} asymmetries~\cite{Hyperon-CP}--\cite{Valencia2000}. Standard-model processes dominate $K$ and $B$ {\em CP} asymmetries, thus it behooves us to study  hyperons (and charm), in which new physics might stand out more sharply. 

More than one hyperon $C\!P$ asymmetry may be measurable in ${\overline p}p$ annihilation. To conserve baryon number, hyperon {\em CP} violation must be of the direct type. Accessible signals include angular-distribution differences of polarized-hyperon and antihyperon  decay products~\cite{ACP}; partial-rate asymmetries, at possibly  detectable levels, are also expected~\cite{Tandean-Valencia,Tandean}. To  compete with previous $\Xi$ and $\Lambda$ {\em CP} studies would require $\sim$\,$10^{33}$ luminosity. While summarizing the state of hyperon {\em CP} violation generally, we therefore emphasize in particular the $\Omega^-/{\overline \Omega}{}^+$ partial-rate asymmetry, for which there is no previous measurement.

By angular-momentum conservation, in the decay of a spin-1/2 hyperon to a
spin-1/2 baryon plus a meson, the final state must be either $S$-wave or $P$-wave.\footnote{A similar argument holds for a spin-3/2 hyperon, but involving $P$ and $D$ waves.}
Interference between the $S$- and $P$-wave decay
amplitudes causes parity violation, described by Lee and
Yang~\cite{Lee-Yang} in terms of two independent parameters $\alpha$ and
$\beta$, proportional respectively to the real and imaginary parts
of the interference term. Hyperon {\em CP}-violation signatures include differences in
$|\alpha|$ or $|\beta|$ between a hyperon decay and its {\em CP}-conjugate
antihyperon decay, as well as particle--antiparticle  decay partial-width differences between a mode and its {\em CP} conjugate~\cite{ACP,Donoghue-etal}. Precision angular-distribution asymmetry measurement requires accurate knowledge of the relative polarizations of the
initial hyperons and antihyperons.

\subsubsection{Angular-distribution asymmetries}

Table~\ref{tab:HCP} summarizes the experimental situation.  
The first three experiments cited studied
$\Lambda$ decay only~\cite{R608}--\cite{PS185} setting limits on the 
{\em CP}-asymmetry parameter~\cite{ACP,Donoghue-etal}
\begin{eqnarray}
A_{\Lambda}\equiv \frac{\alpha_{\Lambda}+
\overline{\alpha}_\Lambda}{\alpha_{\Lambda}-\overline{\alpha}_\Lambda}\,,
\end{eqnarray}
where $\alpha_\Lambda$ ($\overline{\alpha}_\Lambda$) characterizes the
$\Lambda$ ($\overline{\Lambda}$) decay to  (anti)proton plus charged pion. If {\em CP} is a good symmetry in hyperon decay, $\alpha_\Lambda =
-\overline{\alpha}_\Lambda$. 

Fermilab fixed-target experiment E756~\cite{E756} and CLEO~\cite{CLEO} used the decay of charged $\Xi$ hyperons to produce polarized $\Lambda$'s, in whose subsequent decay the
slope of the (anti)proton angular distribution in the ``helicity" frame 
measures the product of $\alpha_\Xi$ and $\alpha_\Lambda$. If {\em
CP} is a good symmetry in hyperon decay this product should be identical for $\Xi^-$ and
$\overline{\Xi}{}^+$ events. The {\em CP}-asymmetry parameter measured is thus 
\begin{eqnarray}
A_{\Xi\Lambda}\equiv \frac{\alpha_{\Xi}\alpha_{\Lambda}-
\overline{\alpha}_\Xi\overline{\alpha}_\Lambda}{\alpha_{\Xi}\alpha_{\Lambda}+
\overline{\alpha}_\Xi\overline{\alpha}_\Lambda}
\approx A_\Xi + A_\Lambda\,.
\end{eqnarray}
%The power of this technique derives from the relatively large $|\alpha|$ value for the $\Xi^-\to\Lambda\pi^-$ decay ($\alpha_\Xi=-0.458\pm0.012$)~\cite{PDG2006}.  A further advantage in the fixed-target case is that within a given ${}^{{}^(\!}\overline{\Xi}{}^{{}^)}$ momentum bin the acceptances and efficiencies for $\Xi^-$ and $\overline{\Xi}{}^+$ decays are very similar, since the switch from detecting $\Xi$ to detecting $\overline{\Xi}$ is made by reversing the polarities of the magnets, making the spatial distributions of decay products across the detector apertures almost identical for $\Xi$ and for $\overline{\Xi}$. (There are still residual systematic uncertainties arising from the differing momentum dependences of the $\Xi$ and $\overline{\Xi}$ cross sections and  of the cross sections for the $p$ and $\overline{p}$ and $\pi^+$ and $\pi^-$ to interact in the material of the spectrometer.)

Subsequent to E756, this technique was used in the ``HyperCP"
experiment (Fermilab E871)~\cite{Holmstrom,Burnstein}, which ran during 1996--99 and has set  the world's best limits on hyperon {\em CP} violation, based so far on about 5\% of the recorded ${}^{{}^(\!}\overline{\Xi}{}^{{}^)}{}^\mp\to{}^{{}^(\!}\overline{\Lambda}{}^{{}^)}\pi^\mp$ data sample.  (The systematics of the full data sample is still under study.) %Like E756, HyperCP used a secondary charged beam produced by 800\,GeV primary protons interacting in a metal target. The secondary beam was  momentum- and sign-selected by means of a curved collimator installed within a 6-m-long dipole magnet. Particle trajectories were measured downstream of a 13-m-long (evacuated) decay region. 
HyperCP recorded the world's largest samples of hyperon and antihyperon decays, including $2.0 \times 10^9$ and $0.46 \times 10^9$  $\Xi^-$ and $\overline{\Xi}{}^+$ events, respectively.   
When the analysis is complete, 
these should determine $A_{\Xi\Lambda}$ with a statistical uncertainty 
\begin{equation}
\delta A = \frac{1}{2\alpha_{\Xi}\alpha_{\Lambda}}
\sqrt{\frac{3}{N_{\Xi^-}}+\frac{3}{N_{\overline{\Xi}{}^+}}} \stackrel{<}{_\sim} 2\times10^{-4}\,.
\label{eq:ACP}
\end{equation} 
The standard model predicts~\cite{ACP} this
asymmetry to be of order $10^{-5}$. (A number of standard-model extensions, e.g.\ nonminimal SUSY, predict effects as large as ${\cal O}(10^{-3})$~\cite{non-SM}.) 
Thus any significant effect seen in HyperCP will be evidence for new sources of {\em CP}
violation in the baryon sector. Such an observation could be of relevance to the mysterious mechanism that gave rise to the  cosmic baryon asymmetry. 

HyperCP has also set the world's first limit on {\em CP} violation in ${}^{{}^(}\overline{\Omega}{}^{{}^)}{}^\mp$ decay, using a sample of 5.46~(1.89)~million $\Omega^-\to\Lambda K^-$  $({\overline\Omega}{}^+\to{\overline\Lambda} K^+)$ events~\cite{Lu-CP}. Here, as shown by HyperCP~\cite{Chen,Lu}, parity is only slightly violated: $\alpha=(1.75\pm0.24)\times10^{-2}$~\cite{PDG2006}. Hence the measured magnitude and uncertainty of the  asymmetry parameter $A_{\Omega\Lambda}$ (inversely proportional to $\alpha$ as in Eq.~\ref{eq:ACP}) are rather large: $[-0.4\pm9.1\,(\rm stat)\pm8.5\,(syst)]\times10^{-2}$~\cite{Lu-CP}. This asymmetry is predicted to be $\le4\times10^{-5}$ in the standard model but can be as large as $8\times10^{-3}$ if new physics contributes~\cite{Tandean}.

\subsubsection{Partial-rate asymmetries}

While {\em CPT} symmetry requires identical lifetimes for particle and antiparticle, partial-rate asymmetries violate only {\em CP}. For most hyperon decays, these are expected to be undetectably small~\cite{Valencia2000}. However, for the decays $\Omega^- \to \Lambda K^-$ and $\Omega^- \to \Xi^0\pi^-$, the particle/antiparticle partial-rate asymmetries could be as large as $2\times10^{-5}$ in the standard model and one to two orders of magnitude larger if non-SM contributions dominate~\cite{Tandean-Valencia,Tandean}. The quantities to be measured are
\begin{eqnarray}\nonumber
\Delta_{\Lambda K} &\equiv&\frac{\Gamma(\Omega^-\to\Lambda K^-)-\Gamma({\overline \Omega}{}^+\to{\overline\Lambda}K^+)}{\Gamma(\Omega^-\to\Lambda K^-)+\Gamma({\overline \Omega}{}^+\to{\overline\Lambda}K^+)}\\
\nonumber &\approx&\frac{1}{2\Gamma}(\Gamma-{\overline \Gamma})
\approx 0.5\,(1-N/{\overline N})
\end{eqnarray}
(and similarly for $\Delta_{\Xi \pi}$),
where in the last step we have assumed nearly equal numbers ($N$) of $\Omega$ and (${\overline N}$) of ${\overline \Omega}$ events, as would be the case in ${\overline p}p$ annihilation. Sensitivity at the $10^{-4}$ level then requires ${\cal O}(10^7)$ reconstructed events. 
Measuring such a small branching-ratio difference reliably will require the clean exclusive ${\overline\Omega}{}^+\Omega^-$ event sample produced less than a $\pi^0$ mass above threshold, or $4.938<p_{\overline p}< 5.437$\,GeV/$c$. 

%\subsubsection{Hyperon sensitivity estimates}

\begin{figure}
\centerline{\hspace{-.05in}\includegraphics[width=\linewidth]{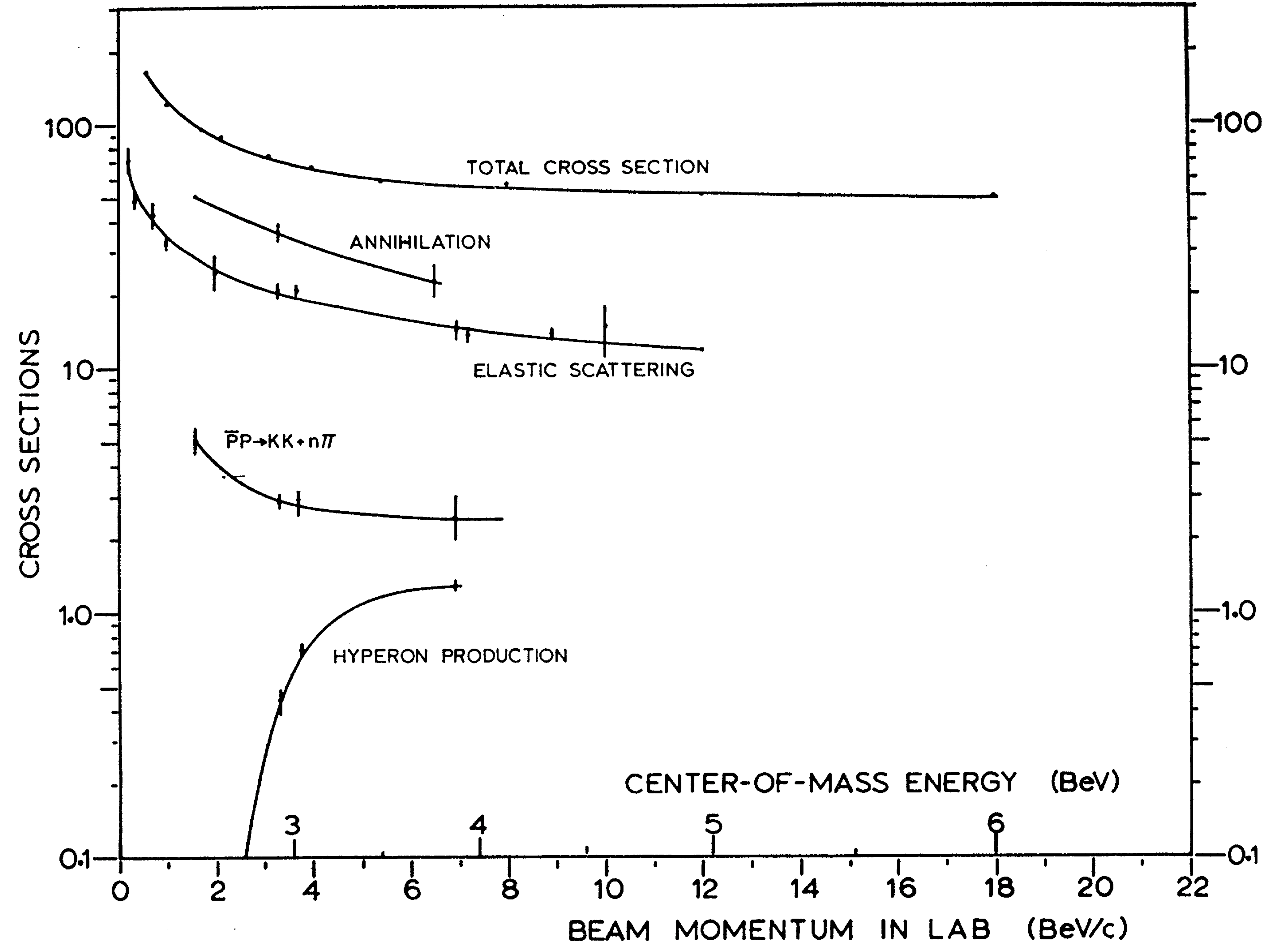}}
\caption{Cross sections (in mb) for various  ${\overline p}p$ processes vs.\ momentum and $\sqrt{s}$ (from \protect\cite{Chien}).}\label{fig:sigma-pbar}
\end{figure}

%There have been a number of measurements of hyperon production by low-energy antiprotons.  
The inclusive hyperon-production cross section at 5.4\,GeV/$c$ is $\approx$\,1\,mb~\cite{pbar-LoI,Kaplan-CTP} (Fig.~\ref{fig:sigma-pbar}).  At $2\times10^{32}\,\rm cm^{-2}s^{-1}$ this amounts to some $2\times10^5$ hyperon events produced per second, or $2\times10^{12}$ per year. (Experience suggests that a data-acquisition system that can cope with such a high event rate is both feasible and reasonable in cost. For example, the ${\overline p}p$ interaction rate is comparable to that in BTeV, yet the charged-particle multiplicity per event is only $\approx$\,1/10 as large.)

To estimate the exclusive ${\overline p}p\to{\overline\Omega}\Omega$ cross section requires some extrapolation, since it has yet to be measured. % (moreover, even for ${\overline p}p\to{\overline \Xi}{}^+\Xi^-$ only a few events have been seen). A rule of thumb is that each strange quark ``costs" between one and two orders of magnitude in cross section, reflecting the effect of the strange-quark mass on the hadronization process.  This is borne out e.g.\ by HyperCP, in which $2.1\times10^9$ $\Xi^-\to\Lambda \pi^-$and $1.5\times10^7$ $\Omega^-\to\Lambda K^-$ decays were reconstructed~\cite{Burnstein}; given the 160\,GeV/$c$ hyperon momentum and 6.3\,m distance from  HyperCP target to decay pipe, this corresponds to $\approx$\,30 $\Xi^-$'s per $\Omega^-$ produced at the target. A similar ratio is observed in HERA-$B$~\cite{Britsch}. In exclusive ${\overline p}p\to{\overline Y}Y$ production (where $Y$ signifies a hyperon) there may be additional effects, since as one proceeds from $\Lambda$ to $\Xi$ to $\Omega$ fewer and fewer valence quarks are in common between the initial and final states. Nevertheless, t
The  cross section for ${\overline \Xi}{}^+\Xi^-$ somewhat above threshold ($p_{\overline p}\approx3.5\,$GeV/$c$) is $\approx$\,2\,$\mu$b~\cite{Hamann,Baltay,HERAG}, or about 1/30 of the corresponding cross section for ${\overline \Lambda}\Lambda$.
Thus the $\approx$\,65\,$\mu$b cross section measured for ${\overline p}p\to{\overline \Lambda}\Lambda$ at $p_{\overline p}=1.642$\,GeV/$c$ at LEAR~\cite{Johansson} implies $\sigma({\overline p}p\to{\overline \Omega}\Omega)\sim60$\,nb at 5.4\,GeV/$c$. 

For purposes of discussion we take this as the exclusive production cross section.\footnote{This estimate will be testable in the upgraded MIPP experiment~\protect\cite{MIPP-upgrade}.} At  $2.0\times10^{32}\,{\rm cm}^{-2}{\rm s}^{-1}$ luminosity, some $1.2\times10^8$ ${\overline \Omega}\Omega$ events are then produced in a nominal 1-year run ($1.0\times10^7$\,s). Assuming 50\% acceptance times efficiency  (comparable to that for $\chi_c$ events in E760), we estimate ${}^{{}^(\!}\overline {N}{}^{{}^)}_{\Xi\pi}=1.4\times10^7$ events each in $\Omega^-\to\Xi^0\pi^-$ and ${\overline \Omega}{}^+\to{\overline\Xi}{}^0\pi^+$, and ${}^{{}^(\!}\overline {N}{}^{{}^)}_{\Lambda K}=4.1\times10^7$ events each in $\Omega^- \to \Lambda K^-$  and ${\overline\Omega}{}^+ \to {\overline\Lambda} K^+$, implying the  partial-rate-asymmetry statistical sensitivities
\begin{eqnarray}
\nonumber \delta\Delta_{\Xi\pi}\approx\frac{0.5}{\sqrt{N_{\Xi\pi}}}\approx1.3\times10^{-4}\,,\\
\nonumber \delta\Delta_{\Lambda K}\approx\frac{0.5}{\sqrt{N_{\Lambda K}}}\approx7.8\times10^{-5}
\,.
\end{eqnarray}
Tandean and Valencia~\cite{Tandean-Valencia} have estimated $\Delta_{\Xi\pi}\approx2\times10^{-5}$ in the standard model but possibly an order of magnitude larger with new-physics contributions.
Tandean~\cite{Tandean} has estimated $\Delta_{\Lambda K}$ to be $\le1\times10^{-5}$ in the standard model but possibly as large as $1\times10^{-3}$ if new physics contributes. (The large sensitivity of $\Delta_{\Lambda K}$ to new physics in this analysis  arises from chromomagnetic penguin operators and final-state interactions via $\Omega\to\Xi\pi\to\Lambda K$~\cite{Tandean}.\footnote{Large final-state interactions  should also affect $\Delta_{\Xi\pi}$ but were not included in that prediction~\protect\cite{Tandean-Valencia,Tandean-private}.}) It is worth noting that these potentially large asymmetries arise from parity-conserving interactions and hence are limited by constraints from  $\epsilon_K$~\cite{Tandean-Valencia,Tandean}; they are independent of $A_\Lambda$ and $A_\Xi$, which arise from the interference of parity-violating and parity-conserving processes~\cite{Tandean-private}. 

Experimental sensitivities will include systematic components whose estimation will require careful and detailed simulation studies yet to be done. Nevertheless, the potential power of the technique is apparent: the experiment discussed here may be capable of observing the effects of new physics in Omega {\em CP} violation via partial-rate asymmetries, and it will  represent a substantial improvement over current sensitivity to Omega angular-distribution asymmetries.

\subsection{Study of FCNC hyperon decays}

\begin{figure}[h]
\centerline{\hspace{-0.3in}\includegraphics[width=.9\linewidth]{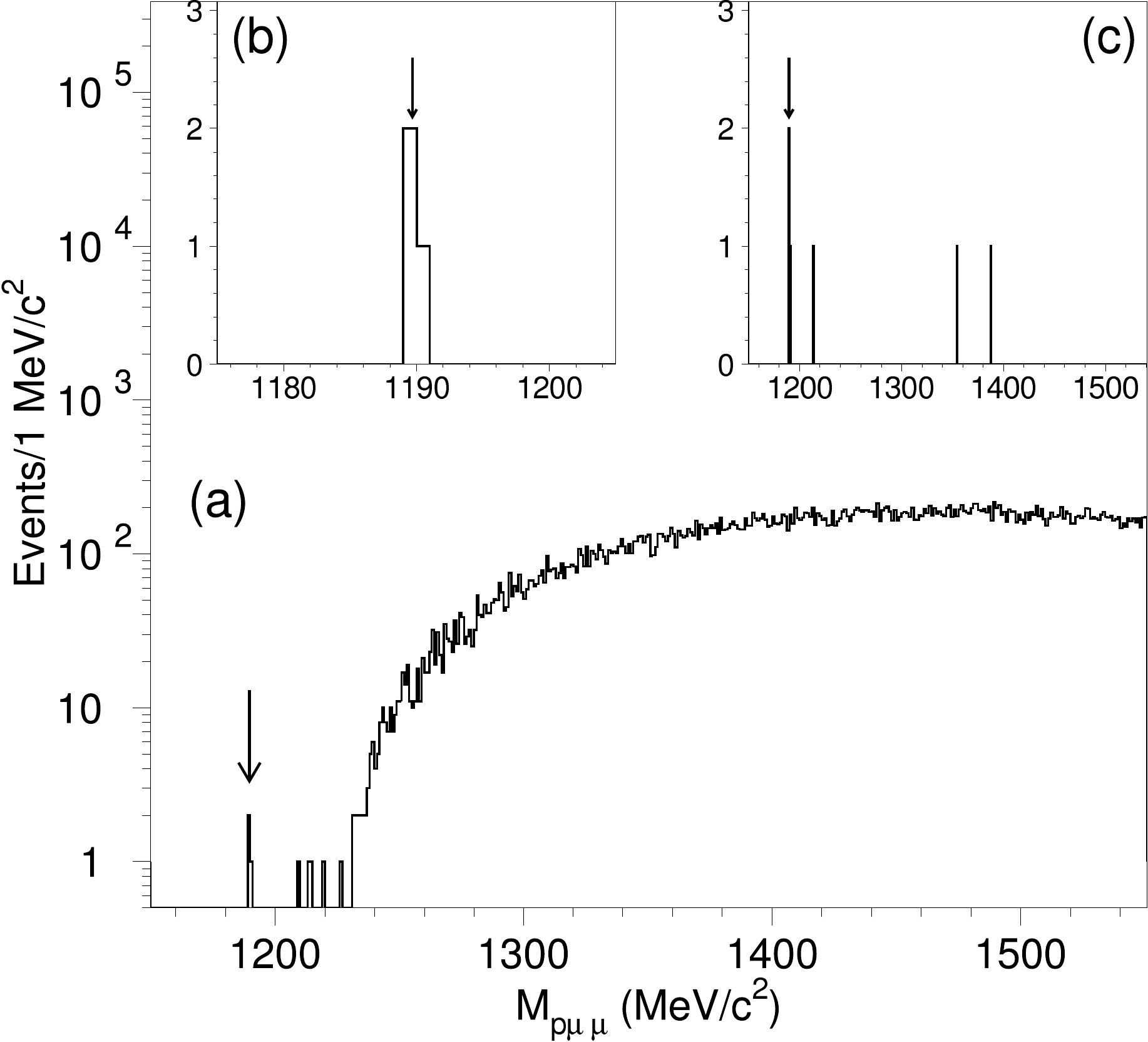}}
\centerline{\hspace{-0.05in}\includegraphics[width=\linewidth]{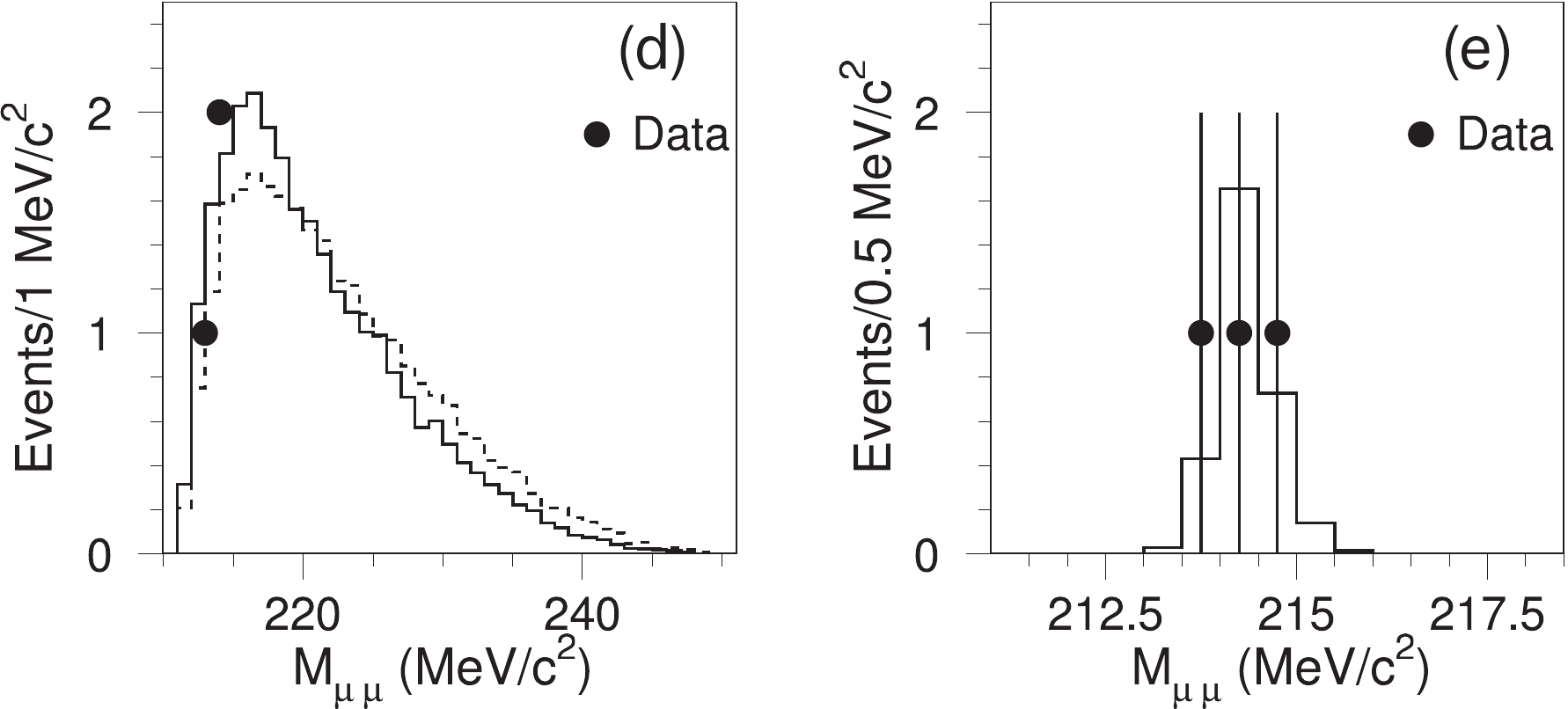}}
\caption{Mass spectra for candidate  single-vertex $p\mu^+\mu^-$  events in HyperCP positive-beam data sample: (a) wide mass range (semilog scale); (b) narrow range around $\Sigma^+$ mass; (c) after application of additional cuts as described in Ref.~\protect\cite{Park-Sigpmumu} (arrows indicate mass of $\Sigma^+$);  dimuon mass spectrum of the
candidate events compared with Monte Carlo spectrum assuming (d) standard-model virtual-photon form factor (solid) or isotropic decay (dashed), or (e) decay via a narrow resonance $X^0$.}
\label{fig:sigpmumu}
\label{fig:mumu}
\end{figure}

Behind its charged-particle spectrometer, HyperCP had muon detectors for  rare-decay studies~\cite{Burnstein,Park-Sigpmumu}.  Using them HyperCP has observed~\cite{Park-Sigpmumu}  the rarest hyperon decay ever, $\Sigma^+\to p\mu^+\mu^-$. Surprisingly (Fig.~\ref{fig:sigpmumu}),  the 3 observed events are consistent with a two-body decay, $\Sigma^+\to p X^0,\,X^0\to\mu^+\mu^-$, with $X^0$ mass $m_{X^0}=214.3\pm0.5\,$MeV/$c^2$. This interpretation is of course not definitive, with the confidence level for the form-factor decay spectrum of Fig.~\ref{fig:mumu}d  estimated at 0.8\%. The measured branching ratio is $[3.1\pm2.4\, (\rm stat)\pm1.5\, (syst)]\times 10^{-8}$ assuming two-body, or $[8.6^{+6.6}_{-5.4}\, (\rm stat)\pm  5.5\,(syst)] \times10^{-8}$ assuming three-body $\Sigma^+$ decay.

The $X^0$, if real, cannot be an ordinary hadron. %This result is intriguing in view of Gorbunov's proposal~\cite{Gorbunov} that  certain nonminimal supersymmetric models include a pair of ``sgoldstinos" (supersymmetric partners of Goldstone fermions), which can be scalar or pseudoscalar and  low in mass. 
A light, scalar or vector particle coupling to hadrons and  muon pairs at the required level is ruled out by its non-observation in kaon decays~\cite{He-etal-Sigpmumu}--\cite{Gengetal}.
However, there are at least two possible supersymmetric  interpretations. It could be a pseudoscalar ``sgoldstino"~\cite{He-etal-Sigpmumu}--\cite{Gengetal} or the light pseudoscalar Higgs boson ($A^0_1$) in the next-to-minimal supersymmetric standard model~\cite{He-Tandean-Valencia}.

Searching for this decay with exclusive ${\overline \Sigma}{}^-\Sigma^+$ events just above threshold would require ${\overline p}$ momentum (see Table~\ref{tab:thresh}) well below that previously achieved by deceleration in the Antiproton Accumulator, as well as very high luminosity to access the ${\cal O}(10^{-8})$ branching ratio. An experimentally less challenging but equally interesting objective is the corresponding FCNC decay of the $\Omega^-$, with  ${\cal O}(10^{-6})$ predicted branching ratio~\cite{He-etal-Sigpmumu} if the $X^0$ is real.\footnote{The standard-model prediction is~\protect\cite{Safadi-Singer} ${\cal B}( \Omega^-\to\Xi^-\mu^+\mu^-)=6.6\times10^{-8}$.}
(The larger  branching ratio reflects the additional phase space available compared to that in $\Sigma^+\to p\mu^+\mu^-$.) As above, assuming $2\times10^{32}$ luminosity and 50\% acceptance times efficiency, 120 or 44 events are predicted in the two cases (pseudoscalar or axial-vector $X^0$) that appear to be viable~\cite{He-etal-Sigpmumu,Deshpande-Eilam-Jiang}:
\begin{eqnarray}
\nonumber{\cal B}(\Omega^-\to\Xi^- X_P\to \Xi^-\mu^+\mu^-)= \qquad\qquad \\
\nonumber ~~~~(2.0^{+1.6}_{-1.2}\pm1.0)\times10^{-6}\,,\\ 
\nonumber{\cal B}(\Omega^-\to\Xi^- X_A\to \Xi^-\mu^+\mu^-)= \qquad\qquad\\(0.73^{+0.56}_{-0.45}\pm \nonumber 0.35)\times10^{-6}\,.
\end{eqnarray}
Given the large inclusive hyperon rates at $\sqrt{s}\approx 3.5\,$GeV, sufficient sensitivity might also be available at that setting to confirm the HyperCP $\Sigma^+\to p\mu^+\mu^-$ results. Alternatively, it is possible that a dedicated run just above $\overline{\Sigma}{}^-\Sigma^+$ threshold may have competitive sensitivity; evaluating this will require a detailed simulation study.

\subsection{Additional physics}\label{addl-phys}

Besides the $X(3872)$, the experiment would be competitive for the charmonium and related states mentioned above. The large hyperon samples could enable precise measurement of hyperon semileptonic and other rare decays. The APEX experiment~\cite{APEX} vacuum tank and pumping system could be reinstalled, enabling substantially increased sensitivity for $\overline p$ lifetime and decay modes. 
There is interest in decelerating further (e.g., at the ends of stores) for trapped-antiproton and antihydrogen experiments~\cite{Holzscheiter,Phillips}.  This capability could make Fermilab the premier facility for such research. The $\overline p$ intensity available at Fermilab could enable  studies not feasible at the AD, such as a measurement of the gravitational force on antimatter~\cite{Phillips}.  A complementary approach is the study of antihydrogen atoms in flight~\cite{Blanford}, which may overcome some of the difficulties encountered in the trapping experiments.

\section{A New Experiment}

We see two approaches to implementing low-cost apparatus to perform the measurements here described~\cite{pbar-LoI}: one based on existing equipment from E835, and the other  on the D\O\ superconducting solenoid (available once the Tevatron Collider program ends). 
Should sufficient resources be available, a new spectrometer, free of constraints from existing apparatus, may give better performance than either of these. The possibility of building a new storage ring has also been mentioned. We hope to study these options in detail in the coming months. 
\section*{Acknowledgments}

This work was supported by the US Dept.\ of Energy under grant DEFG02-94ER40840. The author thanks all of his pbar collaborators and especially D. Christian, K. Gollwitzer, G. Jackson, R. Mussa, C. Patrignani, S. Pordes, J. Rosen, and J. Rosner for useful and stimulating conversations.

\onecolumn{
\begin{table}
\caption {Summary of experimental limits on {\em CP} violation in hyperon decay; the hyperons studied are indicated by $^*$, $^\dagger$, and $^\ddag$.}
\label{tab:HCP}
\begin{center}
\begin{tabular}{lccccc}
\hline\hline
Exp't & Facility & Year & Ref. & Modes & $^*A_\Lambda\,/\,^\dagger A_{\Xi\Lambda}\,/\,^\ddag A_{\Omega\Lambda}$ \\
\hline\hline
R608 & ISR & 1985 & \cite{R608} & $pp\to\Lambda X, pp\to\overline{\Lambda} X$ &
$-0.02\pm0.14^*$ \\
DM2 & Orsay & 1988 &  \cite{DM2} & $e^+e^- \to J/\psi \to \Lambda\overline{\Lambda}$ & 
$0.01\pm0.10^*$
\\
PS185 & LEAR & 1997 & \cite{PS185} & $\overline{p}p\to\overline{\Lambda}\Lambda$ & 
$0.006\pm0.015^*$ \\
& & & &
$e^+e^-\to\Xi^- X, \Xi^-\to\Lambda\pi^-,$ &\\
\raisebox{1.5ex}[0pt]{CLEO} & \raisebox{1.5ex}[0pt]{CESR} &\raisebox{1.5ex}[0pt]{2000} &
\raisebox{1.5ex}[0pt]{\cite{CLEO}} & $e^+e^-\to\overline{\Xi}{}^+ X, \overline{\Xi}{}^+\to\overline{\Lambda}\pi^+$ & 
\raisebox{1.5ex}[0pt]{$-0.057\pm0.064\pm
0.039^\dagger$}\\ 
&  & & &
$pN\to\Xi^- X, \Xi^-\to\Lambda\pi^-$, &  \\
\raisebox{1.5ex}[0pt]{E756} &\raisebox{1.5ex}[0pt]{FNAL} & \raisebox{1.5ex}[0pt]{2000} &
\raisebox{1.5ex}[0pt]{\cite{E756}} & $pN\to\overline{\Xi}{}^+ X, \overline{\Xi}{}^+\to\overline{\Lambda}\pi^+$ & 
\raisebox{1.5ex}[0pt]{$0.012
\pm0.014^\dagger$} \\
&  & & &
$pN\to\Xi^- X, \Xi^-\to\Lambda\pi^-$, &  \\
\raisebox{1.5ex}[0pt]{HyperCP} &
\raisebox{1.5ex}[0pt]{FNAL} & \raisebox{1.5ex}[0pt]{2004} &  \raisebox{1.5ex}[0pt]{\cite{Holmstrom}} &
$pN\to\overline{\Xi}{}^+ X, \overline{\Xi}{}^+\to\overline{\Lambda}\pi^+$ & 
\raisebox{1.5ex}[0pt]{$(0.0
\pm6.7)\times10^{-4}{}^{\,\dagger,\S}$} \\
&  & & &
$pN\to\Omega^- X, \Omega^-\to\Lambda K^-$, &  \\
\raisebox{1.5ex}[0pt]{HyperCP} &
\raisebox{1.5ex}[0pt]{FNAL} & \raisebox{1.5ex}[0pt]{2006} &  \raisebox{1.5ex}[0pt]{\cite{Lu-CP}} &
$pN\to\overline{\Omega}{}^+ X, \overline{\Omega}{}^+\to\overline{\Lambda} K^+$ & 
\raisebox{1.5ex}[0pt]{$-0.004\pm 0.12^{\,\ddag}$} \\
\hline\hline
\end{tabular}
\end{center}
\footnotesize$^\S$ Based on $\approx$5\% of the HyperCP data sample; analysis of the full sample is still in progress.
\end{table}
}
\end{document}